# Deep-ocean inertial subrange small bandwidth coherence and Ozmidov-frequency separation

## by Hans van Haren

NIOZ Royal Netherlands Institute for Sea Research and Utrecht University, P.O. Box 59,
1790 AB Den Burg, the Netherlands.
e-mail: hans.van.haren@nioz.nl

ABSTRACT

The inertial subrange of turbulence in a density stratified environment is the transition from internal waves to isotropic turbulence, but it is unclear how to interpret its extension to anisotropic 'stratified' turbulence. Knowledge about stratified turbulence is relevant for the dispersal of suspended matter in geophysical flows, such as in most of the ocean. For studying internal-wave-induced ocean-turbulence moored high-resolution temperature (T-)sensors are used. Spectra from observations on episodic quasi-convective internal wave breaking above a steep slope of large seamount Josephine in the Northeast-Atlantic demonstrate an inertial subrange that can be separated in two parts: A large-scale part with relatively coherent portions adjacent to less coherent portions, and a small-scale part that is smoothly continuous (to within standard error). The separation is close to the Ozmidov frequency, and coincides with the transition from anisotropic/quasi-deterministic stratified turbulence to isotropic/stochastic inertial convective motions as inferred from a comparison of vertical and horizontal co-spectra. These observations contrast with T-sensor observations of shear-dominated internal wave breaking in an equally turbulent environment above the slope of a small Mid-Atlantic ridge-crest, which demonstrate a stochastic inertial subrange throughout.

I. INTRODUCTION

Turbulence in large Reynolds number geophysical environments like the atmosphere and ocean is often hampered by the stabilizing density stratification that limits the vertical extent of overturns (e.g., Gargett et al., 1984; Davidson, 2013). On the other hand, the associated diapycnal mixing is rather efficient, at least more efficient than in near-homogeneous waters such as can be found in frictional boundary layers adjacent to flat horizontal surfaces. This is because internal waves, which are supported by the stratification, rapidly restratify localized homogeneous patches by their straining and three-dimensional propagation. Despite being important for the dispersal of matter in the ocean, details of the character of 'stratified turbulence' (as turbulence in a stratified environment may be called for short) are not all known. In particular, the transition-range between non-turbulent internal waves and the dissipative Kolmogorov (1941) scales $L_K = (\nu^3/\varepsilon)^{1/4}$ of isotropic turbulence is not often studied in detail from ocean observations. Here $\nu \approx 10^{-6}$ m$^2$ s$^{-1}$ denotes the kinematic viscosity and $\varepsilon$ the turbulence viscous dissipation rate. The transition-range is considered between macroscopic and microscopic scales and the downscale cascade of energy transport between the two is proposed to scale with frequency ($\sigma$) as $\sigma^{-5/3}$ (Obukhov, 1949; Corrsin, 1951; Ozmidov, 1965). The range is loosely termed the inertial subrange (e.g., Tennekes and Lumley, 1972), as it may contain small-scale internal waves supported by thin-layer stratification, anisotropic turbulence affected by the stratification, motions at the Ozmidov scale $L_O = (\varepsilon/N^3)^{1/2}$ of largest isotropic overturn in a stratified environment, besides the full three-dimensional (3D) isotropic turbulence at length-scales $L_K < L < L_O$. N denotes the buoyancy frequency, which separates freely propagating internal waves at its lower frequency and turbulent motions at its higher frequency.

Frehlich et al. (2008) presented velocity spectra from the stable nocturnal atmospheric boundary layer with an inertial subrange that extended over (at least) two orders of magnitude in frequency and which included the Ozmidov frequency $\sigma_O = U/L_O$, where U is a relevant velocity scale. The scaled spectra were not particularly different at frequencies higher and lower than $\sigma_O$. However, while the isotropic motions at $\sigma > \sigma_O$ may be associated with small-scale theory uniform statistics proposed by Kolmogorov (1941), the likely anisotropic motions at $\sigma < \sigma_O$ and the smooth transition between the two regimes without a spectral gap remained unexplained (Davidson, 2013). Riley and Lindborg (2008) proposed to describe the two inertial subranges as 'stratified turbulence' (named by Lilly, 1983) and 'inertial convective subrange' (as in Tennekes and Lumley, 1972), respectively. While Riley and Lindborg's (2008) model exhibits a downscale energy cascade mainly, Lilly (1983) suggested a few percent upscale energy-transfer in the stratified turbulence range that also contains internal waves and quasi-horizontal meandering motions.

Recent modelling results on stratified turbulence, mainly from Direct Numerical Simulations DNS with output commonly in vertical wavenumber ($k_z$) spectra, indicate a sharp transition in spectral scaling, e.g., for the vertical kinetic energy from $k_z$-scaling $k_z^{-3}$ of a buoyancy-inertial subrange to $k_z^{-5/3}$ (e.g., Kimura and Herring, 2012; Augier et al., 2015; Maffioli, 2017). The results confirmed open ocean observations (Gregg, 1977), of which temperature frequency spectra showed an extensive super-buoyancy range scaling with $\sigma^{-3}$ before dropping into $\sigma^{-5/3}$ (van Haren and Gostiaux, 2009). Alisse and Sidi (2000) found indications that the two power scaling-laws were associated with calm and turbulent conditions in the atmosphere. Numerical modelling by Waite (2011) showed the necessity of existence of a distinct stratified turbulence subrange between Ozmidov and buoyancy scales, as large-scale vortices transfer energy to the latter scale via shear instabilities. Kimura and Herring (2012) interpreted the large-scale vortex component to be consistent with $k_z^{-3}$, and the wave component to be consistent with $k_z^{-2}$ the scaling attributed to internal waves (Garrett and Munk, 1972) and fine-structure contamination (Phillips, 1971). All DNS spectra were relatively smooth over one to two orders of magnitude, without small-range variations apart from the broad range changes in power-law scaling (which are changes in slopes on a log-log plot).

In contrast with the atmospheric observations of Frehlich et al. (2008) and with 30-s sampled deep-ocean temperature spectra resolving just the stratified turbulence range (Bouruet-Aubertot et al., 2010), recent deep-ocean high-resolution 1-s sampled temperature spectra from a small-scale five-line 3D mooring array above steep topography demonstrated a two orders of magnitude wide inertial subrange $N < \sigma <$ roll-off with distinctly different small-range variability in the low- and high-frequency parts (van Haren et al., 2016). The distinction was not found in the smooth DNS-spectra (e.g., Augier et al., 2015; Maffioli, 2017). The low-frequency part showed small-range quasi-coherent portions, while the high-frequency part a smooth and well-defined variance-scaling. In co-spectra the two parts were reasonably well defined describing anisotropic and isotropic motions, respectively, and the (continual) transition was associated with twice the smallest (maximum) local buoyancy scale. However, a connection with $L_O$ was not made.

In this paper, the recent 1-s sampled deep-ocean temperature observations, which have about hundred-fold higher resolution than those discussed by Bouruet-Aubertot et al. (2010), are analyzed on the $L_O$-transition and on the quasi-coherent parts. The observations are compared with contrasting ones from a mooring above the slope of a small crest where tidal current shear dominates over (quasi-forced) convection. As ocean turbulence is considered to be largely maintained by internal wave interaction with underwater topography (e.g., Eriksen, 1982; Thorpe, 1987), the focus is on particular internal wave regimes and episodic wave breaking. It is noted that also in the deep ocean flows have high bulk Reynolds numbers Re ~ $O(10^5$-$10^7)$ and that convective and shear-driven overturning occur side-by-side at various scales (e.g., Matsumoto and Hoshino, 2004; Li and Li, 2006). The observations may also be a useful addition to recent advances in laboratory and numerical modelling of flows in confined basins where a boundary is present, still at generally lower Re than in the ocean or atmosphere, for further detailing the demonstrated multiple scale turbulence statistics of natural convection (e.g., Stevens et al., 2011; Augier et al., 2015).

## II. TECHNICAL DETAILS

### A. Temperature sensors

Self-contained moorable high-resolution NIOZ4 T-sensors are used that have been designed for observing internal wave – large-scale turbulence overturns in water (van Haren, 2018). Their noise level is better than $10^{-4}$ °C, their precision is better than $5\times10^{-4}$ °C after correction for instrumental drift. In water, the response time is about 0.5 s, which does not resolve the Kolmogorov dissipation scale, but which resolves the Ozmidov scale and a substantial part of the inertial subrange in most geophysical fluid environments. The T-sensors sample at a rate of 1 Hz. They are synchronized via induction every 4 hours, so that the timing mismatch is <0.02 s.

The calibrated and drift-corrected data are transferred to Conservative (~potential) Temperature ($\Theta$) values using the Gibbs-SeaWater software described in (Intergovernmental Oceanographic Commission (IOC), Scientific Committee on Oceanic Research (SCOR), International Association for the Physical Sciences of the Oceans (IAPSO), 2010). This compensates for the slight, but important in weakly stratified waters, compressibility effects under the large static pressure. Shipborne Conductivity-Temperature-Depth (CTD) profiles near the moored instrumentation are used to evaluate the relative contributions of salinity and temperature to potential density variations. After establishment of a tight temperature-potential density relationship, the T-sensor data can be used as a tracer for potential density to quantify turbulence, as follows.

Turbulence dissipation rate $\varepsilon = c_1^2 d^2 N^3$ is calculated from the T-sensor data using the method of reordering potentially unstable vertical density profiles in statically stable ones, as proposed by Thorpe (1977). Here, d denotes the displacements between unordered (measured) and reordered profiles. N denotes the buoyancy frequency computed from the reordered profiles. In the highly turbulent, stratified, restratifying and relatively strong shear and convection environment over deep-ocean topography, standard constant mean values (over many realizations spread over one order of magnitude) are used of $c_1 = 0.8$ for the Ozmidov/overturn scale factor (Osborn, 1980; Dillon, 1982; Oakey, 1982). Hereafter, averaging over the vertical is indicated by < . . . >, over time by […]. Averaging intervals will be indicated, and the 1-Hz sampling rate ensures a large number of realizations in an average.

**B. Moorings and sites**

Data are analyzed from two T-sensor moorings. A 230 m high mooring was deployed for 10 days at 36° 23.56′N, 33° 53.62′W in 770 m water depth near the crest of an elongated small seamount bounding the axial graben of the Mid-Atlantic Ridge (van Haren et al., 2017). The mooring held 98 T-sensors at 2 m intervals, the lowest at z = 5 m above the seafloor. Three sensors showed calibration problems or too high noise levels and are not further considered. Their data are linearly interpolated. Two single-point Nortek AquaDopp acoustic current meters were attached at z = 6 and 201 m. They sampled at once per 10 s. As verified with pressure and tilt sensors, the top of the mooring did not move more than 0.3 m vertically and it never deflected more than 10 m horizontally, under maximum 0.5 m s$^{-1}$ flow speeds. CTD-data from stations within 1 km from the mooring provided a reasonably tight relationship of $\delta\sigma_{0.65} = \alpha\delta\Theta$, $\alpha$ = -0.15±0.03 kg m$^{-3}$ °C$^{-1}$ over the vertical range of T-sensors. Here, the potential density anomaly $\delta\sigma_{0.65}$ is referenced to 650 dbar.

The second mooring, a small-scale 3D thermistor array, was moored at 37° 00′N, 013° 48′W in 1740 m water depth on a steep slope of the eastern flank of large Mont Josephine, about 400 km southwest of Lisbon (Portugal) in the Northeast-Atlantic (van Haren et al., 2016). The average local bottom slope of about 10º was more than twice steeper (supercritical) than the average slope of semidiurnal internal tides under local stratification conditions. The site was also well below the Mediterranean Sea outflow, between 1000 and 1400 m, so that salinity compensated apparent density inversions in temperature were minimal. The local temperature-potential density anomaly referenced to 1600 dbar ($\sigma_{1.6}$) relationship was $\delta\sigma_{1.6} = \alpha\delta\Theta$, $\alpha$ = -0.044±0.005 kg m$^{-3}$ °C$^{-1}$. The foldable mooring held 475 T-sensors at 1 m vertical intervals, distributed over 5 lines, 105 m tall and 4 m apart horizontally and 104 m long, when fully stretched. The volume sampled was about 3000 m$^3$. The 1000 N (100 kgf) tension on each line was sufficient to have a relatively stiff mooring, with little motion under current drag. Tilt was small (<1°). Heading information showed commonly less than 1° variation of compass data around their mean flow values, except smaller than 10° variations during three brief, relatively strong (~0.22 m s$^{-1}$) flow speed events. The lowest T-sensor was also at z = 5 m. Due to various problems, 33 of the 475 sensors did not function properly. Their data are not considered. Currents were measured using a large-scale resolving 75 kHz acoustic Doppler current profiler at a separate mooring about 1 km away.

The amount of good T-sensors and the 2-m vertical resolution were sufficient to use these moored data to determine turbulence values to within a factor of two through the resolution of scales of up to the largest energy-containing scales of 10-50 m.

## III. OBSERVATIONS AND DISCUSSION

### A. General

At both sites, semidiurnal (lunar) tidal motions dominate flows and mean turbulence values are relatively high with dissipation rates O(10$^{-7}$) m$^2$ s$^{-3}$, which is typical for energetic internal wave breaking above sloping topography and 100 to 1000 times larger than found in the open ocean (Gregg, 1989; Polzin et al., 1997). For best inter-comparison and on computational grounds we analyse from each of the two observational mooring sites one set of 4 days of 0.5 Hz and 2 m vertical interval (sub-)sampled data for the common range between z = 5 and 99 m. The about two times shallower site over the small Mid-Atlantic ridge-crest demonstrates about twice larger dissipation rates, twice larger buoyancy frequency, twice smaller tidal vertical excursion amplitude and about equal tidal current amplitude in comparison with the site over large Mont Josephine. It is noted that the analyzed vertical range does not resolve Mont Josephine's internal (tidal) wave excursion.

Over the analyzed 4 days, 94-m range of T-sensors mean turbulence values are:

For the small Mid-Atlantic ridge-crest mooring: $[<\varepsilon>] = 14\pm8\times10^{-7}$ m$^2$ s$^{-3}$, $[<N>] = 3.8\pm0.9\times10^{-3}$ s$^{-1}$, $[U] = 0.11$ m s$^{-1}$, $[<L_O>] = 5.5$ m, $\sigma_O = 0.02$ s$^{-1}$ (260 cpd, short for cycles per day).

For the large Mont Josephine mooring: $[<\varepsilon>] = 6\pm4\times10^{-7}$ m$^2$ s$^{-3}$, $[<N>] = 1.9\pm0.4\times10^{-3}$ s$^{-1}$, $[U] = 0.11$ m s$^{-1}$, $[<L_O>] = 9$ m, $\sigma_O = 0.011$ s$^{-1}$ (165 cpd).

**B. Small seamount-crest shear-dominated motions**

A typical time-depth image from above the Mid-Atlantic ridge-crest demonstrates a tidal periodicity that is just about resolved in the lower 100 m above the seafloor, and which is superposed with smaller scale internal wave motions of higher frequency (Fig. 1a). The interaction between the two provides a varying modulated signal. In detail (Fig. 1b), shorter scale internal wave motions and shear-induced overturning are visible, especially near the interface around z = 120 m. Below this in the lower 100 m above the seafloor, downward phase propagation is visible with downward draught of turbulent convective overturning around day 184.12.

The spectral information of the above observations suggests a shear-dominated turbulence also for the lower 100 m above the seafloor, because the passive scalar temperature variance spectrum indicates a clear inertial subrange that extends over nearly two orders of magnitude in frequency (Fig. 2b) (Tennekes and Lumley, 1972; Warhaft, 2000). The statistical convergence upon increase of spectral smoothing is uniform through the inertial subrange, as exhibited by the equal vertical extent spread (spectral thickness) between approximately 50 and 5000 cpd, or between N and the spectral roll-off. The vertical extent spread increases somewhat for $\sigma < N$, especially around tidal harmonic frequencies. This is attributable to the peak influence of deterministic signals like tidal harmonics. The main slope is about +2/3 in the log-log plot, or a power-law scaling of $\sigma^{-1}$, which is commensurate with open-ocean internal waves (van Haren and Gostiaux, 2009). This band does not scale with the canonical $\sigma^{-2}$ internal wave scaling (Garrett and Munk, 1972), which would give a slope of -1/3 here. The relatively broad hump around 30 cpd, which is well-known from the open ocean (Munk, 1980) at frequencies just lower than N, is associated with near-buoyancy frequency internal waves that are naturally supported by the main stratification. Its high-frequency flank slopes at approximately -4/3, or a power-law scaling of $\sigma^{-3}$.

This spectral hump is also reflected in vertical current spectra as expected considering the linear wave relationship w $\propto$ $T_t$, and, somewhat unexpected, in horizontal current spectra (Fig. 3). In general, the aspect ratio of vertical over horizontal current variance $|w|^2/E_k$ is (much) smaller than unity for $\sigma < \sigma_O$ in oceanographic data, with a relatively large value of about 0.5 around the near-N peak at 30 cpd. Here, the Ozmidov frequency coincides to within error with the maximum small-scale buoyancy frequency $N_{s,max}$, the maximum buoyancy frequency computed over $\Delta z = 2$ m vertical intervals. In the range $\sigma_O < \sigma < 2\sigma_O$ a transition occurs and the aspect ratio is larger than unity for higher frequencies (probably due to the instrumental configuration of the acoustic beams). This transition reflects the transition of coherence between T-sensors at vertical separation distances of mean Ozmidov scale (5.5 m) and larger from significantly different from zero at lower frequencies to below statistical significance at higher frequencies $>\sim 2\sigma_O$ (Fig 2a). Coherence between T-sensors at vertical scales of 16 m and larger drop rapidly to noise levels for $\sigma > N$, while coherence at the 2 m scale is still (barely) significant close to the roll-off frequency, for all the 48 T-sensors from the lower 100 m above the seafloor involved in the statistics.

(For the 48 T-sensors in the range between z = 100 and 196 m the variance- and co-spectra are essentially similar to the ones of Fig. 2. The upper current meter aspect ratio of unity is found near N. Less KE- and w-variance than at the lower current meter is observed at all frequencies except near the Nyquist frequency, which is noise dominated, and, for KE only, at semidiurnal (and lower) frequencies. This suggests a distinct redistribution from the internal tide, presumably the major internal wave source, to all other frequencies as turbulence increases towards the bottom. In the internal wave band, the aspect ratio between vertical and horizontal motions is much smaller at the upper CM, especially at semidiurnal frequencies, except near N.)

Although the co-spectra are quite smooth in their transition from high values in the internal wave band at $\sigma < N$ through the inertial subrange to spectral roll-off, a few significant variations in coherence are observable, e.g. consistent small peaks at 60, 90 and 105 cpd (Fig. 2a).

**C. Large seamount steep slope shear-convection motions**

Such small-scale coherence variations as a function of frequency are registered more clearly in data from above a steep slope of Mont Josephine (Fig. 4). Probably coincidental sub-peaks are observed at approximately the same frequencies of 60, 90 and 105 cpd, besides at 25 cpd which is close to N in these data. While maximum small-scale buoyancy frequency is almost identical (about 230 cpd) as in Fig. 2, the Ozmidov frequency is 30% lower, which is still similar to within error bounds or the statistical spread around its rms mean value. The larger (9 m) mean Ozmidov scale in these data more or less indicates the drop to insignificant coherence levels in Fig. 4. At about $2\sigma_O$ equal horizontal and vertical separation distances have equal coherence, suggesting dominating isotropic motions. The transition from coherent internal wave motions at $\sigma < N$ to roll-off is distributed over about two-and-a-half orders of magnitude. However, the largest difference with Fig. 2 is in the character of the internal wave and inertial subranges at frequencies in between $N < \sigma <$ roll-off.

In contrast with Fig. 2b, the T-variance in Fig. 4b does not show a bulge at near-buoyancy frequencies $\sigma \sim < N$, a weak tendency for a slope of $+2/3$ and, as in Fig. 2b, a lack of slope $-1/3$. For $\sigma > N$ the approximate inertial subrange is observed not uniform but to be split in two parts: A low-frequency part that has a vertical spread larger than the stochastic 'error' spread and a stochastic high-frequency part that has a vertical spread equivalent to the error spread. The transition between these two parts is around the maximum small-scale buoyancy frequency, or between $\sigma_O < \sigma < 2\sigma_O$. At $\sigma < \sigma_O$ the statistics of the 442 independent T-sensors distributed over all five lines is almost identical to the one of the 87 independent T-sensors of the central line, instead of being decreased by a factor of $(442/87)^{1/2} \approx 5^{1/2}$ for normally distributed statistics, as is observed for $\sigma > 2\sigma_O$. The difference in statistical convergence suggests a relatively small contribution of random signals and a strong deterministic character for the inertial subrange part $N < \sigma < \sigma_O$ that is similar to the internal wave band. However, strong band-smoothing by averaging the contents of neighboring frequency bands demonstrates a tendency to stochastic values and a collapse to the inertial subrange scaling, which distinguishes the inertial subrange from the internal wave band.

To understand the different inertial subrange character of stratified turbulence in Fig. 4 compared with the one in Fig. 2, its time-depth series is investigated in different bands. It is seen that the range of observations does not resolve the semidiurnal internal tide, which has excursions exceeding 100 m in the vertical (Fig. 5), and which carries and promotes the breaking into convective turbulence, see the detail in (Fig. 6a). The character is different from shear-induced overturning as in Fig. 1, given the fact that a substantial part of the buoyancy-Ozmidov range does not follow the $\sigma^{-5/3}$ inertial subrange scaling. However, it is noted that the convection inherently carries small-scale (secondary) shear-instabilities, such as in the modelling by Li and Li (2006). As a result, a final collapse following strong band-smoothing to inertial subrange scaling of shear-induced turbulence of a passive scalar (Ozmidov, 1965; Tennekes and Lumley, 1972; Warhaft, 2000) is not surprising. However, sensor-smoothed quasi-deterministic portions of relatively high and low coherence and relatively high and low T-variance characterize the stratified turbulence part and may be associated with the internal wave convection.

For some understanding in the time-domain, four double-elliptic sharp and phase-preserving bandpass filters are designed (Fig. 4b) that demonstrate various scales (Fig. 6b-e). Most intense motions occur around day 142.22 in the presented example, just before the change from warming to cooling phase. Around this time in the wave phase, motions at all frequencies show largest temperature variability. The internal wave band near-N motions show large vertical coherence scales (Fig. 6b), as expected. Relatively large vertically uniform scales are also observed near the Ozmidov frequency (Fig. 6d) and relatively small vertical scales at lower and higher frequencies (Fig. 6c and 6e).

## IV. GENERAL DISCUSSION AND CONCLUSIONS

The deep-ocean turbulence observations from above strongly sloping topography are not associated with frictional flows, and thus not with a frictional Ekman boundary layer with a typical extent of 10 m above the seafloor for 0.1 m s$^{-1}$ flow speeds. Instead, the turbulence observations are associated with internal wave breaking. The largest energy input is at the semidiurnal lunar tidal frequency, and the O(100 m) amplitude waves slosh back and forth over the sloping seafloor thereby rapidly restratifying the waters making turbulent mixing rather efficient. A large discrepancy is observed between mainly shear-induced overturning with relatively large high-frequency internal wave content near the buoyancy frequency, and shear-convective overturning. While both examples show an Ozmidov frequency that separates the super-buoyancy frequency range in two, with aspect ratio < 1 motions in the range N < σ < σ$_O$ and isotropic aspect ratio of unity motions at σ > σ$_O$, the statistics of the former low-frequency part of the inertial subrange is found different. In the stratified shear flow case, the range N < σ < σ$_O$ is highly stochastic with an uninterrupted continuation to the general stochastic range σ > σ$_O$. In the shear-convective flow case, the range N < σ < σ$_O$ is apparently partially quasi-deterministic or locally coherent, at least at scales < 25 m roughly. This is the stratified turbulence of some interest (e.g., Lilly, 1983; Riley and Lindborg, 2008; Augier et al., 2015).

The convective motions appear quasi-periodic at the high-frequency internal wave scale near the buoyancy frequency, with an association with the particular (warming) phase of the semidiurnal tide, the large-scale 'carrier' internal wave. It seems that the weak acceleration of the internal tidal wave modulates the high-frequency internal wave to become convectively unstable. Or, oblique propagation of the internal tide over the sloping topography may lead to convective instability. For the latter to occur one needs a large slope, with spatial scales exceeding those of the carrier wave. The internal tide has a horizontal scale O(1 km), which may explain why the small ridge-crest site does not exhibit shear-convection, but highly shear-induced turbulence mainly: Its horizontal spatial scales match those of the internal tide (van Haren et al., 2016). The convection is not necessarily horizontally bounded, but the indirect effects of the topography are the wave steepening and breaking, which is expected to vary over the wave's horizontal scales that set a natural boundary. In addition, the layered stratification sets vertical buoyancy and Ozmidov scale boundaries. While shear-dominated flows may influence the stability of stratification at large and small scales, that of shear-convective flows may have a preference for (secondary) shear organization at the small scales and (primary) convection at the large scales, in the Mont Josephine example presented here. It remains to be investigated how much of the latter stratified turbulence energy is transported upscale as suggested by Lilly (1983) and how the particular frequency distribution is organized, if at all.

The present high-resolution observations may shed some light on laboratory/numerical modelling. For high Re-flows in the deep-ocean it was suggested by Gargett (1988) that the small-scale density layering plays an important role, via the small-scale buoyancy frequency, in determining the scale of separation between anisotropic (lower frequencies) and isotropic (higher frequencies) motions. The frequencies at which isotropy is found likely relate with Froude number equal to one (Billant and Chomaz, 2001). Previous oceanographic observations have shown marginal stability across thin stable stratified layers just balancing relatively high destabilizing shear (van Haren et al. 1999). This shear is mainly found at low near-inertial internal wave frequencies. The Mont Josephine data suggest that, in addition to inertial shear organization in thin layers, semidiurnal tides or other large-scale internal waves may also contribute to turbulent exchange via the initiation of convective instabilities that appear intermittently as stratified turbulence, thereby disturbing the otherwise smooth stochastic nature of the inertial subrange.

**ACKNOWLEDGMENTS**

This research was supported in part by NWO, the Netherlands Organization for the advancement of science. I thank the captain and crew of the R/V Pelagia and NIOZ-MTM for their very helpful assistance during deployment and recovery. I thank M. Laan, L. Gostiaux, J. van Heerwaarden, R. Bakker and Y. Witte for all discussions and trials during design and construction of the instrumentation.

**REFERENCES**

Alisse, J. R., and Sidi, C., "Experimental probability density functions of small-scale fluctuations in the stably stratified atmosphere," J. Fluid Mech. 402, 137-162 (2000).

Augier, P., Billant, P., and Chomaz, J.-M., "Stratified turbulence forced with columnar dipoles: numerical study," J. Fluid Mech. 769, 403-443 (2015).

Bendat, J. S., and Piersol, A. G., "Random data: analysis and measurement procedures," John Wiley, New York, 566 pp (1986).

Billant, P., and Chomaz, J.-M., "Self-similarity of strongly stratified inviscid flows," Phys. Fluids 13, 1645-1651 (2001).

Bouruet-Aubertot, P., van Haren, H., and LeLong, M.-P., "Stratified inertial subrange inferred from in situ measurements in the bottom boundary layer of Rockall Channel," J. Phys. Oceanogr. 40, 2401-2417 (2010).

Corrsin, S., "On the spectrum of isotropic temperature fluctuations in an isotropic turbulence," J. Appl. Phys. 22, 469-473 (1951).

Davidson, P. A., "Turbulence in rotating, stratified and electrically conducting fluids," Cambridge University Press, Cambridge UK, 681 pp (2013).

Dillon, T. M., "Vertical overturns: a comparison of Thorpe and Ozmidov length scales," J. Geophys. Res. 87, 9601-9613 (1982).

Eriksen, C. C., "Observations of internal wave reflection off sloping bottoms," J. Geophys. Res. 87, 525-538 (1982).

Frehlich, R., Meillier, Y., and Jensen, M. L., "Measurements of boundary layer profiles with in situ sensors and Doppler lidar," J. Atmos. Ocean. Technol. 25, 1328-1340 (2008).

Garrett, C. J. R., and Munk, W. H., "Space-time scales of internal waves," Geophys. Fluid Dyn. 3, 225-264 (1972).

Gargett, A. E., Osborn, T. R., and Nasmyth, P. W., "Local isotropy and the decay of turbulence in a stratified fluid," J. Fluid Mech. 144, 231-280 (1984).

Gargett, A. E., "The scaling of turbulence in the presence of stable stratification," J. Geophys. Res. 93, 5021-5036 (1988).

Gregg, M. C., "Variations in the intensity of small-scale mixing in the main thermocline," J. Phys. Oceanogr. 7, 436-454 (1977).

Gregg, M. C., "Scaling turbulent dissipation in the thermocline," J. Geophys. Res. 94, 9686-9698 (1989).

IOC, SCOR, and IAPSO, "The international thermodynamic equation of seawater – 2010: Calculation and use of thermodynamic properties," Intergovernmental Oceanographic Commission, Manuals and Guides No. 56, UNESCO, Paris, France, 196 pp (2010).

Kimura, Y., and Herring, J. R., "Energy spectra of stably stratified turbulence," J. Fluid Mech. 698, 19-50 (2012).

Kolmogorov, A. N., "The local structure of turbulence in incompressible viscous fluid for very large Reynolds numbers," Dokl. Akad. Nauk. SSSR 30, 301-305 (1941).

Li, S., and Li, H., "Parallel AMR code for compressible MHD and HD equations," T-7, MS B284, Theoretical division, Los Alamos National Laboratory, https://es.scribd.com/document/74751081/amrmhd (accessed 4 April 2019), (2006).

Lilly, D. K., "Stratified turbulence and the mesoscale variability of the atmosphere," J. Atmos. Sci. 40, 749-761 (1983).

Maffioli, A., "Vertical spectra of stratified turbulence at large horizontal scales," Phys. Rev. Fluids 2, 104802, https://doi.org/10.1103/PhysRevFluids.2.104802 (2017).

Matsumoto, Y., and Hoshino, M., "Onset of turbulence by a Kelvin-Helmholtz vortex," Geophys. Res. Lett. 31, L02807, doi:10.1029/2003GL018195 (2004).

Munk, W. H., "Internal wave spectra at the buoyant and inertial frequencies," J. Phys. Oceanogr. 10, 1718-1728 (1980).

Oakey, N. S., "Determination of the rate of dissipation of turbulent energy from simultaneous temperature and velocity shear microstructure measurements," J. Phys. Oceanogr. 12, 256-271 (1982).

Obukhov, A. M., "Structure of the temperature field in a turbulent flow," Izv. Akad. Nauk. SSSR Ser. Geogr. Geofiz. 13, 58-69 (1949).

Osborn, T. R., "Estimates of the local rate of vertical diffusion from dissipation measurements," J. Phys. Oceanogr. 10, 83-89 (1980).

Ozmidov, R. V., "About some pecularities of the energy spectrum of oceanic turbulence," Dokl. Akad. Nauk SSSR 161, 828-831 (1965).

Phillips, O. M., "On spectra measured in an undulating layered medium," J. Phys. Oceanogr. 1, 1-6 (1971).

Polzin, K. L., Toole, J. M., Ledwell, J. R., and Schmitt, R. W., "Spatial variability of turbulent mixing in the abyssal ocean," Science 276, 93-96 (1997).

Riley, J. J., and Lindborg, E., "Stratified turbulence: A possible interpretation of some geophysical turbulence measurements," J. Atmos. Sci. 65, 2416-2424 (2008).

Stevens, R. J. A. M., Lohse, D., and Verzicco, R., "Prandtl and Rayleigh number dependence of heat transport in high Rayleigh number thermal convection," J. Fluid Mech. 688, 31-43 (2011).

Tennekes, H., and Lumley, J. L., "A first course in Turbulence," MIT Press, Cambridge, 300 pp (1972).

Thorpe, S. A., "Turbulence and mixing in a Scottish loch," Phil. Trans. Roy. Soc. Lond. A 286, 125-181 (1977).

Thorpe, S. A., "Transitional phenomena and the development of turbulence in stratified fluids: a review," J. Geophys. Res. 92, 5231-5248 (1987).

van Haren, H., "Philosophy and application of high-resolution temperature sensors for stratified waters," Sensors 18, 3184, doi:10.3390/s18103184 (2018).

van Haren, H., and Gostiaux, L., "High-resolution open-ocean temperature spectra," J. Geophys. Res. 114, C05005, doi:10.1029/2008JC004967 (2009).

van Haren, H., Maas, L., Zimmerman, J.T.F., Ridderinkhof, H., Malschaert, H., "Strong inertial currents and marginal internal wave stability in the central North Sea," Geophys. Res. Lett. 26, 2993-2996 (1999).

van Haren, H., Cimatoribus, A. A., Cyr, F., and Gostiaux, L., "Insights from a 3-D temperature sensors mooring on stratified ocean turbulence," Geophys. Res. Lett. 43, 4483-4489, doi:10.1002/2016GL068032 (2016).

van Haren, H., Hanz, U., de Stigter, H., Mienis, F., and Duineveld, G., "Internal wave turbulence at a biologically rich Mid-Atlantic seamount," PLoS ONE 12(12), e0189720 (2017).

Waite, M. L., "Stratified turbulence at the buoyancy scale," Phys. Fluids 23, 066602 (2011).

Warhaft, Z., "Passive scalars in turbulent flows," Ann. Rev. Fluid Mech. 32, 203-240 (2000).

**FIG. 1.** High-resolution Conservative Temperature observations above a Mid-Atlantic Ridge crest. (a) Four day time-full 200 m depth-range series. The periods of the inertial frequency (f) and the semidiurnal lunar tidal frequency ($M_2$) are indicated by the black and white horizontal bar, respectively. The detailed period of b. is indicated by the purple bar. The seafloor is at the horizontal axis, 770 m water depth. (b) Magnification (4.8 h) of a. The period of the buoyancy frequency (N) is indicated by the black horizontal bar.

**FIG. 2.** Temperature spectra for the lower 94 m (48 T-sensors) of the vertical range of Fig. 1a. The data are sub-sampled at 0.5 Hz for computational reasons. (a) Average coherence spectra between all pairs of independent sensors for the labelled vertical ($\Delta z$) intervals. The low dashed lines tending to coherence of 0.07 at high frequencies indicate the approximate 95%-significance levels computed following (Bendat and Piersol, 1986). Several buoyancy frequencies are indicated by dashed vertical lines including four-day large-scale mean N and the maximum small-scale (thin-layer) $N_{s,max}$. The mean Ozmidov scale-length is indicated by the cross on the heavy solid line of the Ozmidov-frequency. (b) Corresponding power spectra are scaled with inertial subrange $\sigma^{-5/3}$ (the horizontal black-dashed line on log-log plot. The slope of -1/3 (purple) indicates the canonical internal wave slope (Garrett and Munk, 1972) and finescale structure (Phillips, 1971), while +2/3 (blue) indicates the open-ocean internal wave slope and -4/3 (green) the open-ocean slope for frequencies just higher than N (van Haren and Gostiaux, 2009). In light-blue the 48-sensor average T-spectrum, weakly band-smoothed. In red, the same spectrum more heavily band-smoothed, in blue strongly band-smoothed and slightly shifted vertically for clarity. Inertial, semidiurnal lunar tidal and Ozmidov ($\sigma_O$) frequencies are indicated. The 95% significance levels are indicated by the small vertical bars for spectra of corresponding colour.

**FIG. 3.** Variance spectra as Fig. 2b but from lower current meter data with kinetic energy in black and vertical component in purple.

**FIG. 4.** As Fig. 2, but for four days of 3D-mooring data from Mont Josephine. In (a) also horizontal separation distances are indicated for the two thicker-line coherence spectra. In (b) the light-blue spectrum is for 87-sensor independent records from the single (central) line only, the also weakly smoothed red spectrum is for all 442 independent records from the 5 lines, the blue spectrum its strongly band-smoothed version. In black are spectra of four different band-pass filters.

**FIG. 5.** As Fig. 1a, but for four days of 3D-mooring central line data from a slope of Mont Josephine, 1740 m water depth. The purple line indicates the detailed period of Fig. 6.

**FIG. 6.** The upper panel shows a 4.8 h magnification from Fig. 5, the lower panels its different band-pass filtered versions from low- to high-frequencies. (For filter bounds see black spectra in Fig. 4b).

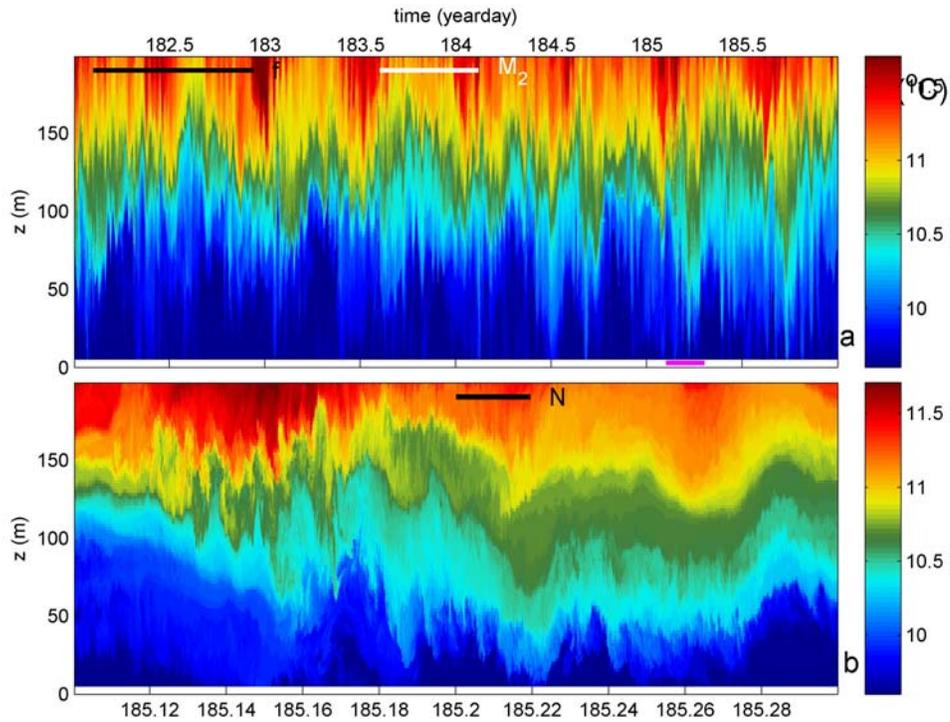

**FIG. 1.** High-resolution Conservative Temperature observations above a Mid-Atlantic Ridge crest. (a) Four day time-full 200 m depth-range series. The periods of the inertial frequency (f) and the semidiurnal lunar tidal frequency ($M_2$) are indicated by the black and white horizontal bar, respectively. The detailed period of b. is indicated by the purple bar. The seafloor is at the horizontal axis, 770 m water depth. (b) Magnification (4.8 h) of a. The period of the buoyancy frequency (N) is indicated by the black horizontal bar.

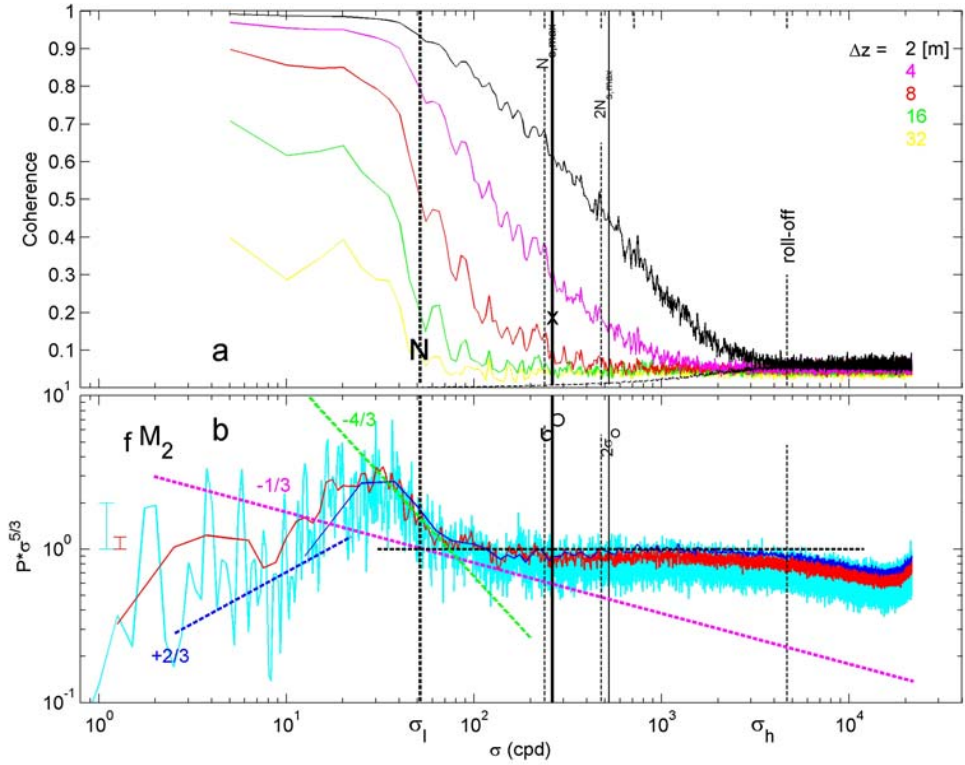

**FIG. 2.** Temperature spectra for the lower 94 m (48 T-sensors) of the vertical range of Fig. 1a. The data are sub-sampled at 0.5 Hz for computational reasons. (a) Average coherence spectra between all pairs of independent sensors for the labelled vertical ($\Delta z$) intervals. The low dashed lines tending to coherence of 0.07 at high frequencies indicate the approximate 95%-significance levels computed following (Bendat and Piersol, 1986). Several buoyancy frequencies are indicated by dashed vertical lines including four-day large-scale mean N and the maximum small-scale (thin-layer) $N_{s,max}$. The mean Ozmidov scale-length is indicated by the cross on the heavy solid line of the Ozmidov-frequency. (b) Corresponding power spectra are scaled with inertial subrange $\sigma^{-5/3}$ (the horizontal black-dashed line on log-log plot. The slope of -1/3 (purple) indicates the canonical internal wave slope (Garrett and Munk, 1972) and finescale structure (Phillips, 1971), while +2/3 (blue) indicates the open-ocean internal wave slope and -4/3 (green) the open-ocean slope for frequencies just higher than N (van Haren and Gostiaux, 2009). In light-blue the 48-sensor average T-spectrum, weakly band-smoothed. In red, the same spectrum more heavily band-smoothed, in blue strongly band-smoothed and slightly shifted vertically for clarity. Inertial, semidiurnal lunar tidal and Ozmidov ($\sigma_O$) frequencies are indicated. The 95% significance levels are indicated by the small vertical bars for spectra of corresponding colour.

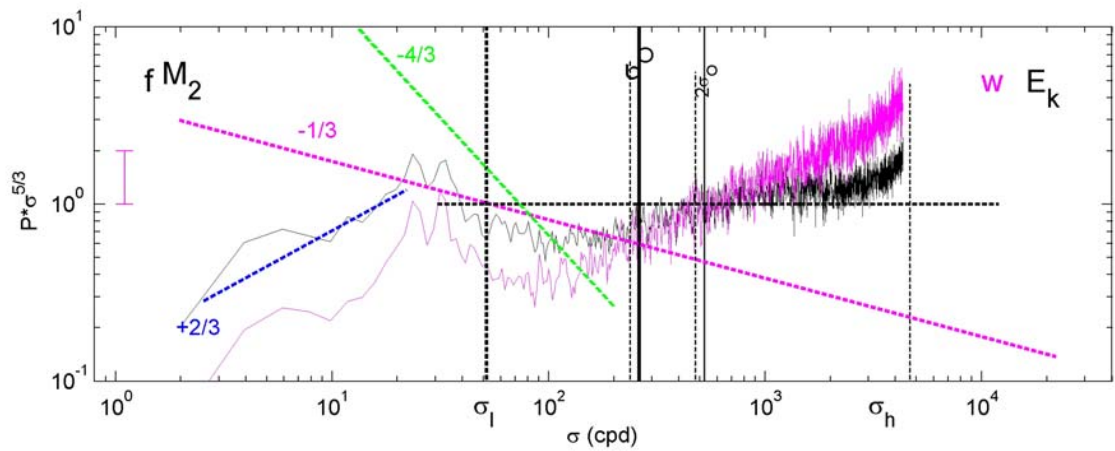

**FIG. 3.** Variance spectra as Fig. 2b but from lower current meter data with kinetic energy in black and vertical component in purple.

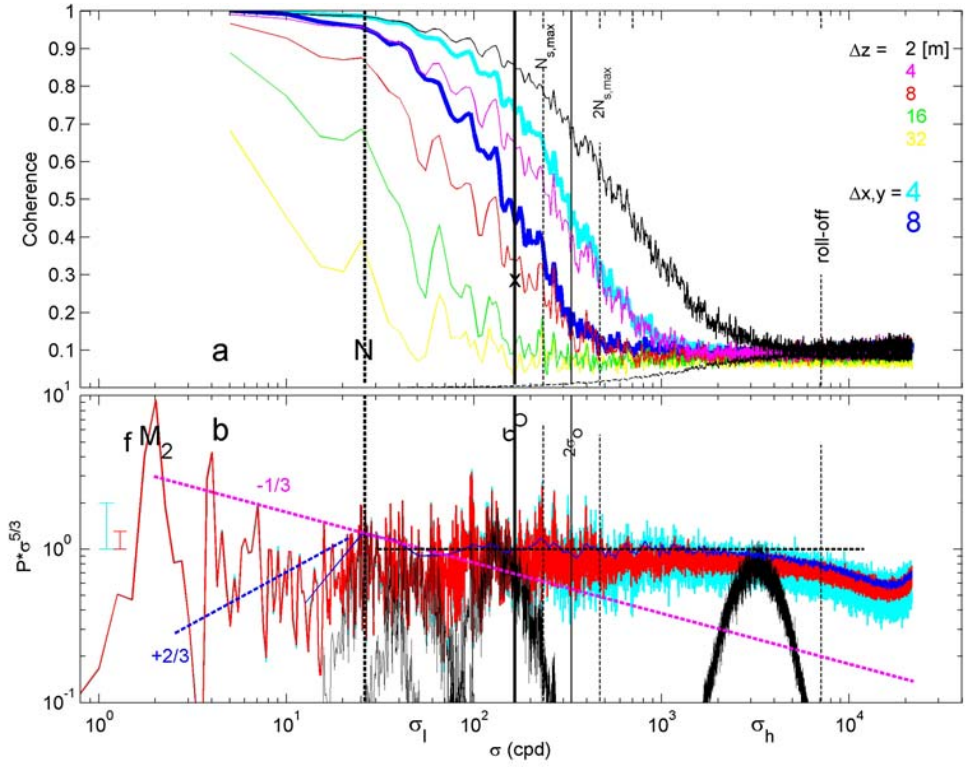

**FIG. 4.** As Fig. 2, but for four days of 3D-mooring data from Mont Josephine. In (a) also horizontal separation distances are indicated for the two thicker-line coherence spectra. In (b) the light-blue spectrum is for 87-sensor independent records from the single (central) line only, the also weakly smoothed red spectrum is for all 442 independent records from the 5 lines, the blue spectrum its strongly band-smoothed version. In black are spectra of four different band-pass filters.

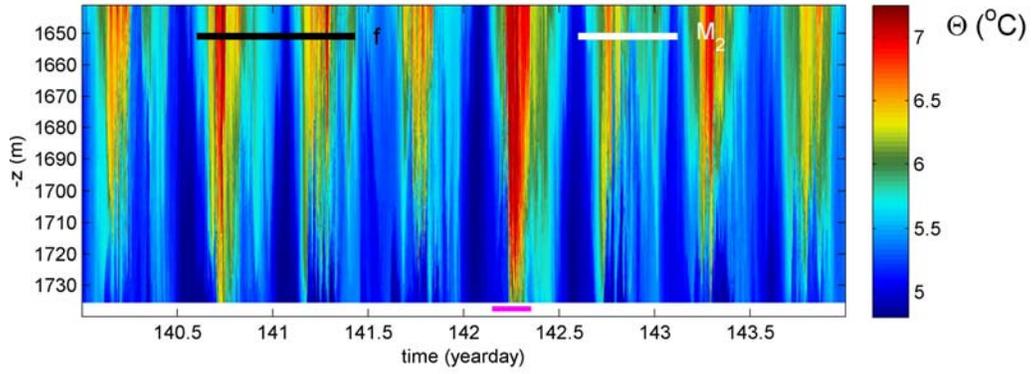

**FIG. 5.** As Fig. 1a, but for four days of 3D-mooring central line data from a slope of Mont Josephine, 1740 m water depth. The purple line indicates the detailed period of Fig. 6.

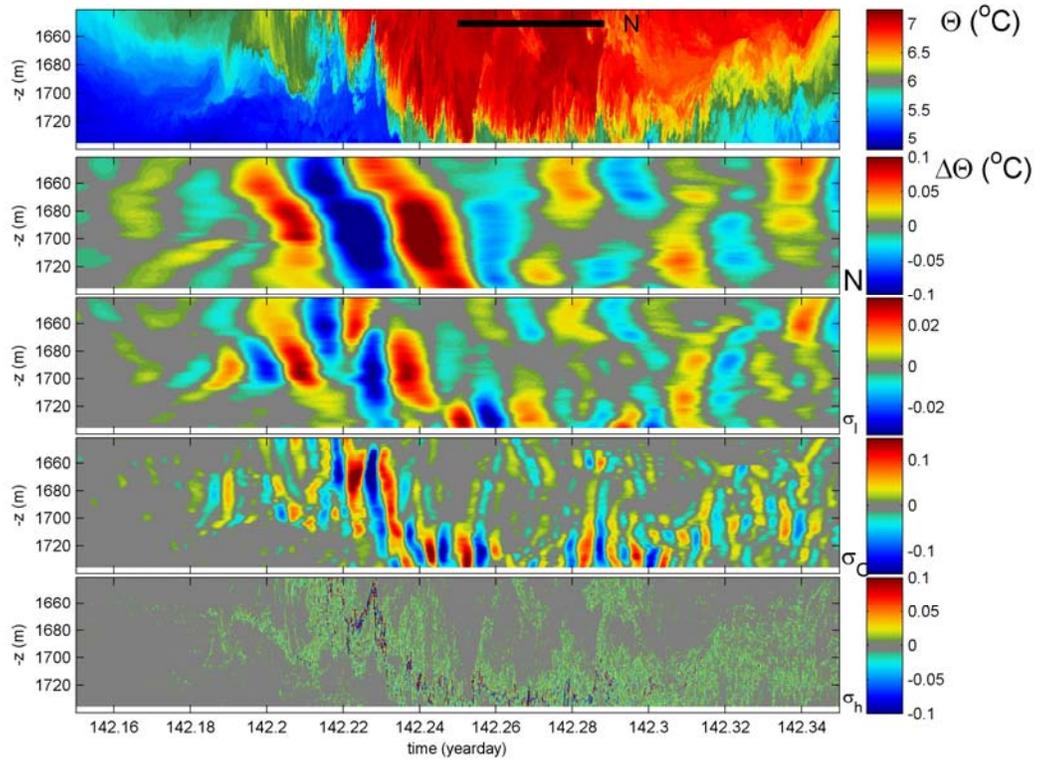

**FIG. 6.** The upper panel shows a 4.8 h magnification from Fig. 5, the lower panels its different band-pass filtered versions from low- to high-frequencies. (For filter bounds see black spectra in Fig. 4b).

27